# Voltage Dependence of Spin Transfer Torque in Magnetic Tunnel Junctions

M. Chshiev[1], I. Theodonis[2,3], A. Kalitsov[2,4], N. Kioussis[2], and W. H. Butler[1], *Member, IEEE*

[1]Center for Materials for Information Technology, University of Alabama, Tuscaloosa, AL 35487, USA
[2]Department of Physics and Astronomy, California State University, Northridge, CA 91330, USA
[3]Department of Physics, National Technical University, Zografou, GR-15773, Greece
[4]Institute Neel, 38042 Grenoble, France

**Theoretical investigations of spin transfer torque in magnetic tunnel junctions using the tight-binding model in the framework of non-equilibrium Green functions formalism are presented. We show that the behavior of the spin transfer torque as a function of applied voltage can vary over a wide range depending on the band parameters of the ferromagnetic electrodes and the insulator that comprise the magnetic tunnel junction. The behavior of both the parallel and perpendicular components of the spin torque is addressed. This behavior is explained in terms of the spin and charge current dependence and on the interplay between evanescent states in the insulator and the Fermi surfaces of ferromagnetic electrodes comprising the junction. The origin of the perpendicular (field-like) component of spin transfer torque at zero bias, i.e. exchange coupling through the barrier between ferromagnetic electrodes is discussed.**

## I. INTRODUCTION

Current induced magnetization switching using spin transfer torque (STT) continues to generate interest for spin electronic applications such as MRAM, spin torque oscillators and detectors [1], [2]. Among the most favorable candidates for realization of STT devices are epitaxial magnetic tunnel junctions (MTJ) [3], [4]. Thus, understanding the fundamental mechanisms that can affect the dependence of Tunneling Magnetoresistance (TMR) and STT on the applied voltage in MTJs is critically important [5]. Unlike fully metal-based nanostructures, at finite applied voltage, the total perpendicular (field like) component of the STT ($T_\perp$) in MTJs is not negligible [6] and in the ballistic regime exhibits quadratic behavior as a function of applied voltage [7]. We recently predicted that the parallel (Slonczeswki) component of STT ($T_\parallel$) could behave non-monotonically as a function of applied voltage [7]. These predictions were recently confirmed experimentally [8], [9].

Here we provide a systematic study of the influence of majority and minority band filling on the applied voltage dependence of both the parallel ($T_\parallel$) and perpendicular ($T_\perp$) terms of STT in MTJs.

## II. MODEL

The calculations have been performed within the tight-binding model using the non-equilibrium Green function technique in the framework of the Keldysh formalism [10], [11]. The magnetic tunnel junction under consideration consists of two ferromagnetic electrodes (FM and FM') separated by an insulator (Fig.1). The magnetizations **M** and **M'** of the electrodes are non-collinear and form an angle $\gamma$ in x-z plane. **M'** defines the quantization axis along the z-axis.

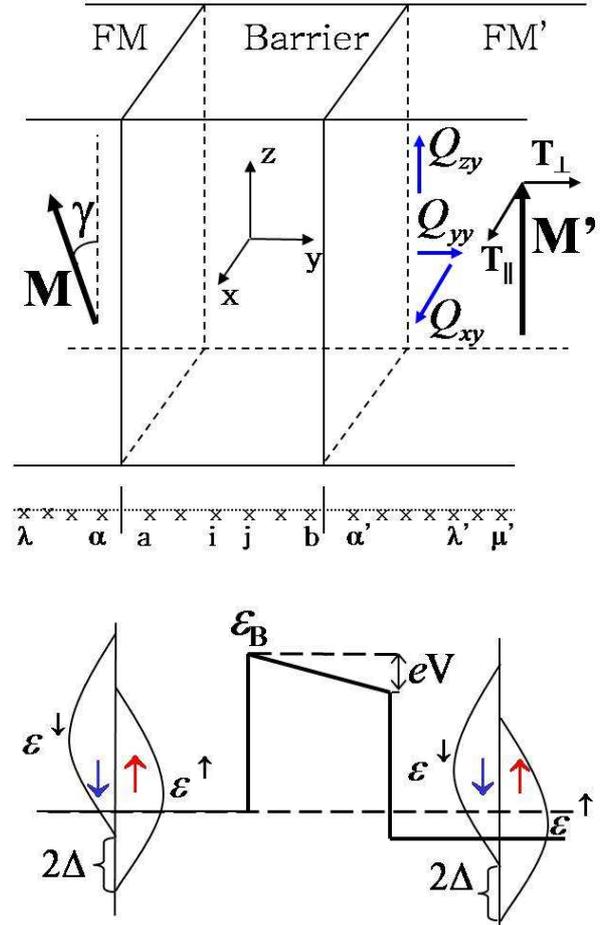

Fig. 1. Schematic representation and corresponding potential profile of magnetic tunnel junction under consideration.

Assuming a linear potential profile across the tunnel junction under applied voltage V, using the notations shown in Fig. 1, and taking into account only the nearest neighbor interactions, the tight-binding Hamiltonian of this junction can

$$H = H_L + H_R + H_B + H_{LB} + H_{RB}$$

$$H_L^\sigma = \sum_\lambda \varepsilon^\sigma c_\lambda^+ c_\lambda + t \sum_{NN} c_\lambda^+ c_\mu$$

$$H_{L(R)} = \frac{H_{L(R)}^\uparrow + H_{L(R)}^\downarrow}{2}\begin{pmatrix}1 & 0 \\ 0 & 1\end{pmatrix} + \frac{H_{L(R)}^\uparrow - H_{L(R)}^\downarrow}{2}\begin{pmatrix}\cos\gamma & \sin\gamma \\ \sin\gamma & -\cos\gamma\end{pmatrix}$$

$$H_R^\sigma = \sum_{\lambda'}(\varepsilon^\sigma + eV)c_{\lambda'}^+ c_{\lambda'} + t\sum_{NN} c_{\lambda'}^+ c_{\mu'}$$

$$H_{LB} = t_{a\alpha}c_\alpha^+ c_a \begin{pmatrix}1 & 0 \\ 0 & 1\end{pmatrix} + h.c.$$

$$H_{RB} = t_{b\alpha'}c_{\alpha'}^+ c_b \begin{pmatrix}1 & 0 \\ 0 & 1\end{pmatrix} + h.c.$$

$$H_B = (\sum_i \varepsilon_B^i c_i^+ c_i + t_B \sum_{NN} c_i^+ c_j)\begin{pmatrix}1 & 0 \\ 0 & 1\end{pmatrix}, \quad \varepsilon_B^i = \varepsilon_B + \frac{i}{N}eV$$

be written using electron creation and annihilation operators $c$ and $c^\dagger$ in the form shown above. Here $\varepsilon^\sigma$ and $t$ represent spin-dependent on-site energies ($\sigma=\uparrow,\downarrow$), and nearest neighbor hopping matrix elements in the FM electrodes, respectively. The exchange splitting $\Delta$ within FM electrodes is defined as $(\varepsilon^\downarrow - \varepsilon^\uparrow)/2$ (see Fig. 1). Similarly, $\varepsilon_B$ and $t_B$ are on-site energies and hopping matrix elements for the insulating barrier whereas $t_{a\alpha}$ and $t_{b\alpha'}$ indicate hopping parameters for the left and right interfaces, respectively and $N$ is the number of layers in the insulator. We consider a simple cubic lattice and take into account translational invariance in $x$-$z$ plane by integrating over the corresponding in-plane wave vector $k_\parallel$.

Using the technique developed in Refs. [10] and [11], we first find the isolated retarded Green functions $g_{\lambda\mu}$, $g_{ij}$ and $g_{\lambda'\mu'}$ corresponding to three regions of the structure where Greek indices denote left and right FM layer sites while Latin ones are used for sites in the barrier (see Fig. 1). Next the retarded Green functions $G_{pq}$ of the whole system are found using the Dyson equation, where $p$ and $q$ may belong to any of three regions of the junction. Finally, the "lesser" non-equilibrium Green function $G_{pq}^<$ is found by solving the kinetic equation [11]. $G_{pq}^<$ is a 2x2 matrix for each $(p,q)$. Let us note here that the latter can be explicitly associated with the electrons coming from the left and right electrode reservoirs far from interfaces, i.e. it can be written as $G_{pq}^< = G_{pq}^{<L} + G_{pq}^{<R}$ where the latter are proportional to Fermi-Dirac distribution functions $f_L(E)$ and $f_R(E\pm eV)$, respectively. Having calculated the $G_{pq}^<$, we can write down the spin current density tensor in the form:

$$\mathbf{Q}_{\lambda'} = \frac{1}{4\pi}\int dE \left[\mathbf{Q}_{\lambda'}^L(E) + \mathbf{Q}_{\lambda'}^R(E)\right] \quad (1)$$

with

$$\mathbf{Q}_{\lambda'}^{L(R)}(E) = \frac{t}{4\pi^2}\int dk_\parallel Tr\left\{\left[G_{\lambda'+1,\lambda'}^{<L(R)\sigma\sigma'} - G_{\lambda',\lambda'+1}^{<L(R)\sigma\sigma'}\right]\boldsymbol{\sigma}\right\} \quad (2)$$

where $\boldsymbol{\sigma}$ represents the vector of Pauli matrices and the trace is taken over spin indices. Because the current direction is along the y-axis, only three components of the spin current tensor, namely $Q_{xy}, Q_{yy}$ and $Q_{zy}$ survive (schematically shown in Fig. 1). Contracting the real space part of the spin current tensor (1)-(2) using the divergence operator [12] along the current direction gives the local torque vector in the right FM electrode $\mathbf{T}_{\lambda'} = \mathbf{Q}_{\lambda'-1} - \mathbf{Q}_{\lambda'}$, where only transverse components, $T_x$ and $T_y$ (or $T_\parallel$ and $T_\perp$), of spin transfer torque survive since the longitudinal spin current $Q_z$ is conserved unlike the transverse ones $Q_x$ and $Q_y$ (see Fig. 1). The total deposited torque in the right FM electrode is then found by summation over all layers [7]:

$$\mathbf{T} = \sum_{\lambda'=0}^{\infty}\mathbf{T}_{\lambda'} = \sum_{\lambda'=0}^{\infty}\mathbf{Q}_{\lambda'-1} - \mathbf{Q}_{\lambda'} = \mathbf{Q}_{-1} - \mathbf{Q}_{\infty} = \mathbf{Q}_{-1} \quad (3)$$

Where we took into account that transverse spin current vanishes far from interface. Thus, in case of semi-infinite FM electrodes in the ballistic regime the total parallel and perpendicular torque components $T_\parallel$ and $T_\perp$ (called sometimes Slonczewski's and field-like term, respectively), are simply equal to the value of the transverse spin current inside the barrier at the FM/insulator interface [7].

The charge current and the longitudinal component of the spin current are conserved across the junction. The current density is obtained by replacing $\boldsymbol{\sigma}/2$ in Eq. (2) by the unit vector along the current direction $y$ times $e/\eta$ [13]:

$$\mathbf{J} = \frac{et}{8\pi^3\eta}\int dE\int dk_\parallel Tr\left\{\left[G_{\lambda'+1,\lambda'}^{<\sigma\sigma'} - G_{\lambda',\lambda'+1}^{<\sigma\sigma'}\right]\right\}\hat{\mathbf{y}} \quad (4)$$

In the following sections we use the above expressions to demonstrate the origin of spin transfer torques both at zero and at finite applied voltage. For simplicity we put all hopping parameters $t = t_{a\alpha} = t_{b\alpha'} = t_B = -1$eV. The calculations presented here are for room temperature, for barrier thickness of three monolayers and for $\varepsilon_B = 9$eV. The corresponding band width for both spin channels is therefore equal to $12t$. The angle between magnetizations is $\pi/2$.

### III. RESULTS AND DISCUSSION

In order to understand the details of the quantum origin of spin transfer torque we start by evaluating the total contributions to the transverse spin currents coming from the left and right FM reservoirs according to (2). The results are shown in Fig. 2. Note that all electrons with energies between the bottom of the majority or minority bands and the Fermi level (plus a tail from temperature smearing above the latter) contribute to the total currents from two opposite directions. When the junction is in equilibrium (Fig. 2, solid lines), these contributions have exactly the same magnitude for parallel components of the spin current density $Q_\parallel^L$ and $Q_\parallel^R$, but due to opposite directions they differ in sign. Therefore, according to (1) and (3), the "net" spin current vanishes as well as the total parallel torque. The situation is completely different for the perpendicular spin current as shown by solid lines in the bottom panel in Fig. 2. Both $Q_\perp^L$ and $Q_\perp^R$ are now exactly the same including the sign which leads to a finite value of the perpendicular (field-like) torque $T_\perp$ even at zero voltage. This zero-voltage torque is, in fact, the exchange coupling between left and right FM electrodes [14], and has been observed experimentally [15].

When a positive (negative) voltage is applied, the parallel total spin currents are changed but the same symmetry as for zero bias still holds for the electrons up to the Fermi level of the left (right) electrode so that the "net" spin current (and the total parallel torque) is simply defined by electrons in the energy "window" between two Fermi levels, i.e. $f_L(E_F)$ and $f_R(E_F \pm eV)$. This is illustrated by dashed and dotted lines in Fig. 2 (top panel).

The effect of applied voltage on perpendicular total spin currents is more interesting (see bottom panel in Fig. 2). Unlike the previous case for which both $Q_\parallel^{L(R)}$ below the lower of two Fermi levels were changing simultaneously, i.e. $Q_\parallel^L(E) = -Q_\parallel^R(E)$ for $E < \min\{f_L, f_R\}$, here left (right) total spin current $Q_\perp^{L(R)}$ decreases (increases) under positive (negative) applied voltage. One can notice, however, the interesting property: $Q_\perp^{L(R)}(E)\big|_{V>0} = Q_\perp^{R(L)}(E - eV)\big|_{V<0}$, which leads to a very important relation for the perpendicular "net" spin current density between positive and negative voltage: $Q_\perp(E)\big|_{V>0} = Q_\perp(E - eV)\big|_{V<0}$. Since the integral in (1) is taken over the whole energy range with help of (3), we come to the conclusion that the total field-like term of the spin transfer torque $T_\perp$ is an even parity function of applied voltage [7].

This is indeed what we observe in Fig. 3 where we present a systematic study of the voltage dependence of the field-like torque $T_\perp$ as a function of majority and minority band filling. In order to cover as wide as possible range of band filling (including the effect of exchange splitting) on $T_\perp$, we fixed the majority on-site energy at $\varepsilon^\uparrow=+3$, 0 and -3eV which corresponds to ¼, ½ and ¾ filling (Fig. 3(a), (b) and (c), respectively). For every case we shift gradually the minority band upwards decreasing its filling. We note that at zero voltage $T_\perp$ is finite describing exchange and all curves are indeed exhibit quadratic (which is an even parity) behavior as a function of applied voltage, i.e. we can write $T_\perp(V) = T_\perp^0 + T_\perp^1 V^2$ where $T_\perp^0$ describes the exchange coupling between FM electrodes. This kind of behavior was observed in exciting recent experiments reported in [8] and [9]. Note that $T_\perp^0$ is a very sensitive to the band structure parameters of the FM electrodes and may be both positive and negative in sign.

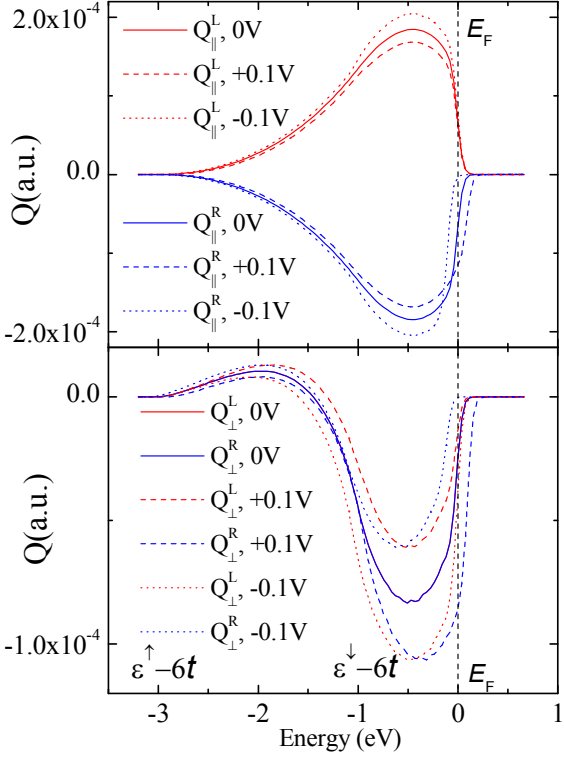

Fig. 2. Energy integrand of the parallel (top panel) and perpendicular (bottom panel) spin current density associated with electron contribution coming from left and right reservoirs. Parameters used: $\varepsilon^\uparrow=3eV$ and $\varepsilon^\downarrow=5eV$.

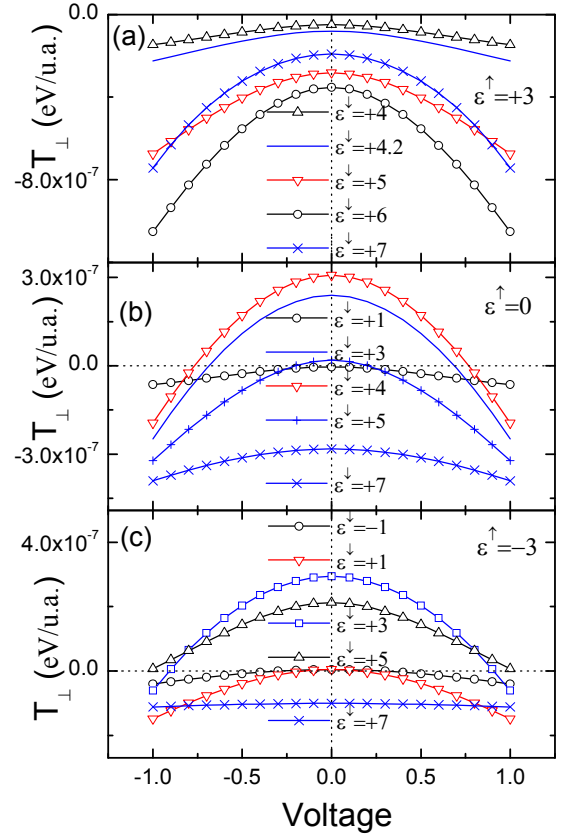

Fig. 3. Total perpendicular (field-like) spin transfer torque per interfacial unit area (u.a.) as a function of applied voltage for different majority and minority band fillings. See text for details.

In Fig. 4 we present the voltage behavior of the parallel component of spin transfer torque exactly for the same band

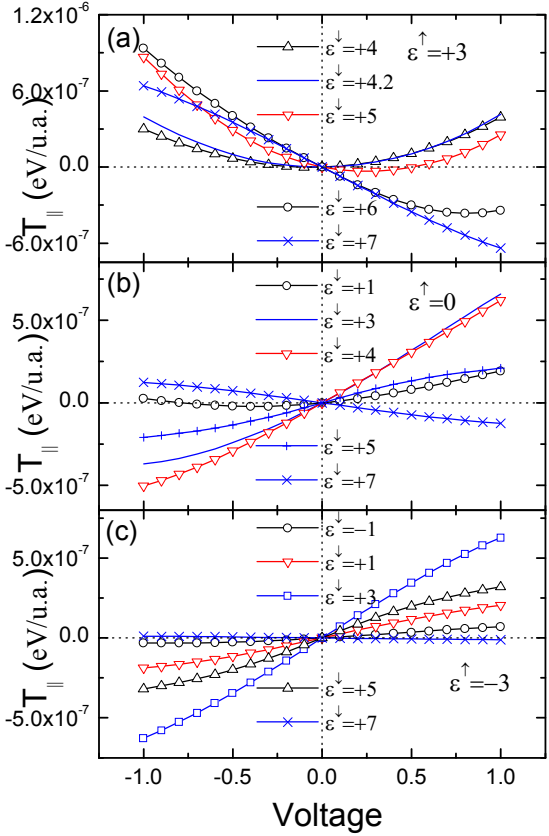

Fig. 4. Total parallel (Slonczweski's) spin transfer torque per interfacial unit area (u.a.) as a function of applied voltage for different majority and minority band fillings. See text for details.

filling cases as in Fig. 3. Unlike the perpendicular torque, the parallel one $T_\parallel$ shows a wide range of non-trivial applied voltage behavior as a function of majority and minority band filling. For instance, for the case of ¼ majority band filling (Fig.4(a)) the parallel component of the spin torque exhibits a wide range of behavior from being a quadratic even function of bias at $\varepsilon^\downarrow$=4.2eV to an odd one for the pure half-metallic case with $\varepsilon^\downarrow$=7eV. The mechanism responsible for such a behavior of parallel component of spin transfer torque can be understood with the generalized equivalent circuit model for MTJ [5], [7]. According to this model the parallel component of the spin transfer torque can be expressed in terms of collinear longitudinal spin currents as

$$T_\parallel(\gamma) = \frac{1}{2}[Q_z(0) - Q_z(\pi)]\mathbf{M}'\times(\mathbf{M}\times\mathbf{M}') \qquad (5)$$

where $Q_z(\gamma) = \eta(J^\uparrow(\gamma) - J^\downarrow(\gamma))/4e$ with charge current density matrix elements from (4). It can be shown that $Q_z(\pi)$ is always an even parity (quadratic) function of applied bias whereas $Q_z(0)$ is an odd parity function of applied voltage [7], [11]. It is clear that in the case of half-metallic electrodes anti-parallel charge as well as spin currents vanish and $T_\parallel$ is defined only by $Q_z(0)$ being an odd parity function of applied voltage as is observed for $\varepsilon^\downarrow$=7eV in Fig. 4(a). More interesting is another limiting case when $Q_z(0)$ vanishes completely yielding the pure quadratic behavior defined only by $Q_z(\pi)$ as in case of $\varepsilon^\downarrow$=4.2eV (Fig. 4(a)). Such a situation is possible due to a maximum in charge current as a function of band filling (or on-site energy) so that the appropriate exchange splitting between majority and minority bands around the maximum causes the corresponding charge currents to be equal [11]. The existence of this maximum can be explained in terms of the charge current dependence on the interplay between evanescent states in the insulator and the Fermi surfaces of the ferromagnetic electrodes comprising the junction. The nature of such interplay is described using the following expression for transmission probability [11]:

$$D = \frac{4\sinh^2 q_y a \sin^2 k_y a}{(\cosh q_y a - \cos k_y a)^2}\exp(-2Nq_y a) \qquad (6)$$

where
$$\cos k_y a = -\varepsilon/2t - \cos k_x a - \cos k_z a$$
$$\cosh q_y a = -\varepsilon_B/2t - \cos k_x a - \cos k_z a$$

Equation (6) is a tight-binding analog of the generalized Julliere model in the free electron case and can be viewed as the product of two identical interfacial transmission probabilities with an exponential decay factor[5], [16], [17]. More detailed analysis will be given in [11].

Thus, the behavior shown in Fig. 4(a) is a result of the interplay between two collinear longitudinal spin currents with different parity properties. Again, recent experiments reported in [8] and [9] show such a behavior. For the case presented in Figs. 4(b) and 4(c) the magnitude of $T_\parallel$ decreases and it is impossible to obtain a situation for which $I^s(0)$ vanishes, moreover, $T_\parallel$ is mostly defined by $I^s(0)$ and the odd parity voltage dependence dominates over even parity. It is interesting to note that in the half-metallic situation, the parallel torquance ($dT_\parallel/dV$) is negative for ¼ majority filling (Fig. 4(a)) and changes sign in case of ½ and ¾ filling (Figs. 4(b) and (c)).

## IV. Conclusion

We provided insights into the quantum origin of both parallel and field-like components of the spin transfer torque for the case of ballistic transport regime in magnetic tunnel junctions. They originate from interplay between total spin current contributions originating from left and right ferromagnetic reservoirs. The parallel and perpendicular total spin currents have different symmetry properties which defines the nature of voltage behavior of Slonczewski's and field-like terms of spin transfer torque and exchange coupling between FM electrodes across insulating barrier. The field-like torque is even parity function of applied voltage whereas Slonczewski's torque may exhibit a wide range of non-monotonic behavior varying from being a purely even to purely odd parity function of applied voltage for the half-metallic case. This is explained by the interplay between evanescent states in the insulator and the Fermi surfaces of the ferromagnetic electrodes comprising the junction. Furthermore, the majority and minority band filling has a dramatic impact on the behavior of spin transfer torque and exchange coupling.

### Acknowledgment

This work was supported by NSF MRSEC Grant N$^o$ DMR 0213985, by the INSIC EHDR Program and was conducted in part at the Center for Nanophase Materials Sciences, which is sponsored at Oak Ridge National Laboratory by the Division of Scientific User Facilities, U. S. Department of Energy.